\documentclass{ws-ijmpd}

\usepackage[super,sort&compress,merge,comma]{natbib}

\usepackage{amsfonts}
\usepackage{amsmath}
\usepackage{amssymb}
\usepackage{mathrsfs}
\usepackage{mathtools}
\usepackage[colorlinks]{hyperref}
\usepackage[hyphenbreaks]{breakurl}

\usepackage{etoolbox}
\usepackage{graphicx}
\usepackage{color}%

\definecolor{darkmidnightblue}{rgb}{0.0, 0.2, 0.4}
\definecolor{darkpowderblue}{rgb}{0.0, 0.2, 0.6}
\definecolor{darkslateblue}{rgb}{0.28, 0.24, 0.55}
\definecolor{magentanew}{rgb}{0.4, 0., 0.4}
\hypersetup{breaklinks=true,colorlinks=true,linkcolor=darkmidnightblue,citecolor=darkmidnightblue,filecolor=darkmidnightblue,urlcolor=darkmidnightblue}
\makeatletter
\patchcmd{\NAT@citex}
  {\@citea\NAT@hyper@{
\NAT@nmfmt{\NAT@nm}
\hyper@natlinkbreak{\NAT@aysep\NAT@spacechar}{\@citeb\@extra@b@citeb}
\NAT@date}}
  {\@citea\NAT@nmfmt{\NAT@nm}
\NAT@aysep\NAT@spacechar\NAT@hyper@{\NAT@date}}{}{}
\patchcmd{\NAT@citex}
  {\@citea\NAT@hyper@{
\NAT@nmfmt{\NAT@nm}
\hyper@natlinkbreak{\NAT@spacechar\NAT@@open\if*#1*\else#1\NAT@spacechar\fi}
{\@citeb\@extra@b@citeb}
\NAT@date}}
  {\@citea\NAT@nmfmt{\NAT@nm}
\NAT@spacechar\NAT@@open\if*#1*\else#1\NAT@spacechar\fi\NAT@hyper@{\NAT@date}}
  {}{}

\def\aj{Astron.~J.}
\def\araa{Ann.~Rev.~Astron.~Astrophys.}
\def\apj{Astrophys.~J.}
\def\apjl{Astrophys.~J.~Lett.}

\def\aap{Astron.~Astrophys.}

\def\mnras{Mon.~Not.~Roy.~Astron.~Soc.}

\def\prd{Phys.~Rev.~D}
\def\jhep{JHEP}

\def\mpla{Mod.~Phys.~Lett.~A}

\def\cqgra{Class.~Quant.~Grav. }
\def\gregr{Gen.~Rel.~Grav.}

\def\epjh{Eur.~Phys.~J.~H}

\def\cmaph{Commun.~Math.~Phys.}

\def\prl{Phys.~Rev.~Lett.}
\def\rvmp{Rev.~Mod.~Phys.}

\def\pasj{Publ.~Astron.~Soc.~Jap.}

\def\ssr{Space~Sci.~Rev.}

\def\nat{Nature}

\def\jpha{J.~Phys.~A: Math.~Gen.}

\def\frp{Front.in Phys.}
\def\ijmpd{Int.~J.~Mod.~Phys.~D}

\begin{document}

\markboth{A. Danehkar}
{Gravitational Fields of the Magnetic-type}

%
\catchline{}{}{}{}{}
%

\title{Gravitational Fields of the Magnetic-type\footnote{This essay received an Honorable Mention in the 2020 Essay Competition of the Gravity Research Foundation.}  }

\author{A. Danehkar}

\address{Department of Astronomy, University of Michigan, \\1085 S. University Avenue, 311 WH, Ann Arbor, MI 48109, USA\\
danehkar@umich.edu}

\maketitle

\begin{history}
\received{15 May 2020}
\accepted{18 June 2020}
\end{history}

\begin{abstract}
Local conformal symmetry introduces the conformal curvature (Weyl tensor)
that gets split into its (gravito-) electric and magnetic (tensor) parts. Newtonian
tidal forces are expected from the gravitoelectric field, whereas
general-relativistic frame-dragging effects emerge from the gravitomagnetic
field. The symmetric, traceless gravitoelectric and gravitomagnetic tensor fields
can be visualized by their eigenvectors and eigenvalues. In this essay, we
depict the gravitoelectric and gravitomagnetic fields around a slowly
rotating black hole. This suggests that the phenomenon of ultra-fast outflows
observed at the centers of active galaxies may give evidence for the
gravitomagnetic fields of spinning supermassive black holes. We also question
whether the current issues in our contemporary observations might be resolved
by the inclusion of gravitomagnetism on large scales in a perturbed FLRW model.
\end{abstract}

\keywords{gravitomagnetism; Weyl tensor; general relativity}

\ccode{PACS numbers: 04.20.$-$q, 95.30.Sf}


~\\
The applicability and validity of Newtonian gravity and classical cosmology have been challenged in both
the weak-gravity limit on large scales and the strong-gravity
regime near supermassive black holes (SMBH). In particular, our observations of Type
Ia supernovae up to the redshift $z \sim 2$ suggested the accelerating expansion of the universe, 
\cite{Riess1998,*Schmidt1998,*Perlmutter1999} which was interpreted as dark
energy.
\cite{Perlmutter1999a,*Peebles2003,*Peebles2003,*Copeland2006,*Riess2007} 
Meanwhile, the rotational velocity curves of visible stars in \textit{disc} (spiral) galaxies
are inconsistent with Kepler's laws of planetary motion,
\cite{Rubin1970,*Rubin1978,*Rubin1980,*Sofue2001} which were explained by
cold dark matter halos enveloping galactic discs.
\cite{Blumenthal1984,*Blumenthal1986,*Kent1987,*Persic1996} Moreover, our
contemporary high-energy observations suggested the presence of
\textit{ultra-fast outflows} with nearly relativistic velocities originated from somewhere close to
SMBHs in \textit{active} galaxies and quasars.
\cite{Tombesi2010,*Tombesi2011,*Tombesi2012,*Kriss2018,*Danehkar2018,*Boissay-Malaquin2019}
 Recently, our understanding of the universe has been revolutionized by the
discovery of gravitational waves resulting from a merger of binary
stellar-mass black holes \cite{Abbott2016,*Abbott2016a} and binary neutron,
stars \cite{Abbott2017,*Abbott2017a} which were predicted by the theory
of general relativity in 1916.\cite{Einstein1916,*Einstein1918,*Einstein1937} 
In general relativity, we also had the prediction of a non-Newtonian field
that is called the \textit{gravitomagetic} field \cite{Thorne1982,*Thorne1986}
by analogy with the magnetic field in Maxwell's theory of electromagnetism.
The Lense--Thirring frame-dragging effect
\cite{Lense1918,*Thirring1918,*Thirring1918a,*Thirring1921} that is one of
the footprints of \textit{gravitomagetism} has been recently detected in a
fast-rotating white dwarf in a binary system.\cite{Krishnan2020} This effect
was previously measured around the Earth using two artificial satellites.
\cite{Ciufolini2004} 

Conformal invariance of Maxwell's equations in electromagnetism has inspired us to explore 
conformal transformations in other fundamental forces of the nature. Considering a local
conformal (Weyl) transformation of the metric, $g_{ab}\rightarrow\Omega
^{2}g_{ab}$ (where $\Omega^{2}$ is the position-dependent conformal factor), we had the introduction of the \textit{Weyl conformal tensor}
$C_{abcd}$ to the Riemann curvature $R_{abcd}$.
\cite{Weyl1918,*Jordan1960,*Jordan2009,*Jordan1961,*Jordan2013,*Ehlers1961,*Ehlers1993,*Kundt1962,*Kundt2016,*Hawking1975} 
The Weyl tensor $C_{abcd}$ 
is conformally invariant and has only 10 independent components. 
A conformal theory of gravity (conformal Weyl gravity) was prescribed 
by an action given by the square of the Weyl tensor \cite{Mannheim1989,*Mannheim1990,*Mannheim1992,*Mannheim1994,*Mannheim1997,*Mannheim2001,*Mannheim2012}  that seems to be spontaneously broken (similar to the
BEH mechanism) in some energy scales, leading to the Einstein-Hilbert action for the
Einstein field equations.\cite{Adler1982,*Hooft2015,*Hooft2017} The Weyl tensor can be split into its electric and magnetic parts, 
i.e. the \textit{gravitoelectric tensor} field $E_{ab}\equiv c^{2}%
C_{acbd}(u^{c}/c)(u^{d}/c)$ and the \textit{gravitomagetic tensor} field
$H_{ab}\equiv-\frac{1}{2}\epsilon_{aecd}C^{cd}{}_{bf}(u^{e}/c)(u^{f}/c)$,
where $u^{a}/c$ is the normalized timelike vector field (such that $u^{a}%
u_{a}=-c^{2}$), $\epsilon_{abcd}$ is the spacetime permutation tensor, 
and $c$ is the speed of light in vacuum. 
The gravitoelectric and gravitomagetic fields are the spatial symmetric, traceless
tensors ($E_{ab}=E_{ba}$, $H_{ab}=H_{ba}$, and $E_{a}{}^{a}=0=H_{a}{}^{a}$), and each has 5 independent components. Equations of motion
for the gravitoelectric and gravitomagetic fields are obtained by substituting
the Einstein field equations into the Bianchi identities.
\cite{Truemper1964,*Hawking1966,*Hawking2014,*Ellis1971,*Ellis1973,*Ellis2009,*Maartens1998,*Ellis1999,*Bertschinger1994,*Danehkar2009}
 Let us consider a perfect fluid model with $T_{ab}=(\rho c^{2}+p)(u_{a}/c)(u_{b}%
/c)+p\eta_{ab}$ that is commonly employed in almost-FLRW spacetimes, where $\eta
_{ab}={\mathrm{diag}}(-c^{2},+1,+1,+1)$ is the Minkowski metric, 
$\rho c^{2}$ is the energy density ($\rho$ is the volumetric mass density),
and $p$ is the isotropic pressure.
For a non-expanding non-accelerated shearless model in a locally almost flat coordinate system, these equations of motion for
$E_{ab}$ and $H_{ab}$ become
\begin{equation}
\mathrm{{D}}^{b}E_{ab}-3c\omega^{b}H_{ab}=\dfrac{8\pi G}{3}\mathrm{{D}}%
_{a}\rho,\label{eq1}%
\end{equation}%
\begin{equation}
\mathrm{{D}}^{b}H_{ab}+\frac{3}{c^{3}}\omega^{b}E_{ab}=-\frac{8\pi G}{c^{3}%
}\omega_{a}(\rho+p/c^{2}),\label{eq2}%
\end{equation}%
\begin{equation}
\operatorname{curl}(E)_{ab}=-c\frac{dH_{ab}}{dt}-cH_{c(a}\omega_{b)}{}%
^{c},\label{eq3}%
\end{equation}%
\begin{equation}
\operatorname{curl}(H)_{ab}=\frac{1}{c^{3}}\frac{dE_{ab}}{dt}+\frac{1}{c^{3}%
}E_{c(a}\omega_{b)}{}^{c},\label{eq4}%
\end{equation}
where $G$ is the gravitational constant, 
$t$ is the time coordinate, $\mathrm{{D}}_{a} \equiv \partial / \partial x^a$ denotes the spatial derivative
with respect to the space coordinates $x^a$, 
$\operatorname{curl}(S)_{ab}%
\equiv\epsilon_{cd(a}\mathrm{{D}}^{c}S_{b)}{}^{d}$\ denotes the spatial curl
of 2nd-rank spatial symmetric tensors, $\omega_{ab}\equiv\mathrm{{D}}%
_{[a}u_{b]}$ is the vorticity tensor, and $u_{b}$ is the velocity vector,
$\omega_{a}\equiv-\frac{1}{2}\epsilon_{abc}\mathrm{{D}}^{b}u^{c}$ is the
vorticity vector, and $\epsilon_{abc} \equiv \epsilon_{abcd} (u^d/c)$ is the spatial permutation tensor. The
round brackets enclosing indices denotes symmetrization (e.g. $A_{(ab)}%
\equiv\frac{1}{2}A_{ab}+\frac{1}{2}A_{ba}$), whereas the square brackets
enclosing indices denotes antisymmetrization (e.g. $A_{[ab]}\equiv\frac{1}%
{2}A_{ab}-\frac{1}{2}A_{ba}$).

Newtonian tidal forces are produced by the gravitoelectric field $E_{ab}$, 
while frame-dragging effects are generated by the gravitomagnetic field $H_{ab}$. 
In the first two equations (\ref{eq1}) and (\ref{eq2}), 
the spatial gradient of the mass density, ($8\pi G/3)\mathrm{{D}}_{a}\rho$,
and the angular momentum density, $-(8\pi G/c^{3})\omega_{a}\left(
\rho+p/c^{2}\right)  $, appear as matter sources for the gravitoelectric and
gravitomagetic fields, respectively. 
In the Newtonian limit ($\mathrm{{D}}^{b}E_{ab}=(8\pi G/3)\mathrm{{D}}_{a}\rho$ and $H_{ab}=0$),
taking $E_{ab} = \mathrm{{D}}_a \mathrm{{D}}_b \Phi - \frac{1}{3} h_{ab} \mathrm{{D}}^2 \Phi$ leads 
to Poisson's equation of Newtonian gravity $\mathrm{{D}}^2 \Phi = 4\pi G \rho$, 
where $\mathrm{{D}}^2 \equiv \mathrm{{D}}_a \mathrm{{D}}^a$
is the Laplace operator, and $h_{ab} ={\mathrm{diag}}(+1,+1,+1)$ is the spatial flat metric. 
The later two equations
(\ref{eq3}) and (\ref{eq4}) support the wave solutions for the gravitoelectric
and gravitomagetic fields, i.e. $\mathrm{D}^{2}E_{ab}-(1/c^{2})d^{2}%
E_{ab}/dt^{2}=0$ and $\mathrm{D}^{2}H_{ab}-(1/c^{2})d^{2}H_{ab}/dt^{2}=0$ 
(see also Ref.~\citenum{Matte1953,*Bel1962,*Bel2000,*DeWitt1962,*Hawking1966b})
in vacuum where the vorticity and matter fields vanish.

We notice an invariance in Eqs.~(\ref{eq1})--(\ref{eq4}) between the
gravitoelectric and gravitomagnetic tensors, $\left(  E_{ab}/c^{2}%
,H_{ab}\right)  \rightarrow\left(  -H_{ab},E_{ab}/c^{2}\right)  $, as well as
the mass density spatial gradient and the angular momentum density, $\left(
\tfrac{1}{3}\mathrm{{D}}_{a}\rho,- \omega_{a}\left(  \rho+p/c^{2}\right)
/c\right)  \rightarrow\left(  \omega_{a}\left(  \rho+p/c^{2}\right)
/c,\tfrac{1}{3}\mathrm{{D}}_{a}\rho\right)  $, which demonstrate a type of the
SO(2) electric-magnetic duality.
\cite{Hull2000,*Hull2001,*Bekaert2004a,*Bunster2013b,*Henneaux2016,*Danehkar2019} The
gravitoelectric and gravitomagnetic fields are transformed into each other
under the electric-magnetic duality rotations, which are analogous with the
electric-magnetic invariance in Maxwell's equations. However, the angular momentum
density appears as a source for the gravitomagnetic field $H_{ab}$ in 
general relativity, whereas we have no magnetic charge for the
magnetic field $\vec{H}$ in Maxwell's theory of electromagnetism.

The gravitoelectric tensor fields $E_{ab}$ generated around a massive object
with the mass quantity are comparable to the electric vector fields $\vec{E}$
around a charged particle with the charge quantity. The gravitomagentic tensor
fields $H_{ab}$ produced around a rotating massive object having the angular
momentum may be compared with the magnetic vector fields $\vec{H}$ around a
bar magnet having the magnetic dipole moment. Nevertheless, we have the 2nd-rank
symmetric traceless tensor fields in gravity rather than the vector (1st-rank
tensor) fields in electromagnetism. We can visualize the physical lines
and amplitudes of the tensor fields by obtaining their eigenvectors and
eigenvalues.\cite{Weickert2006,*Laidlaw2009,*Telea2014} Accordingly, the physical proprieties of $E_{ab}$ and $H_{ab}$
have been visualized based on integral curves of their eigenvectors,
the so-called \textit{tendex} and \textit{vortex} lines, respectively.
\cite{Owen2011,*Nichols2011,*Nichols2012,*Thorne2012,Zhang2012}

\begin{figure*}
\begin{center}
\includegraphics[width=0.75\textwidth, trim = 0 0 0 0, clip, angle=0]{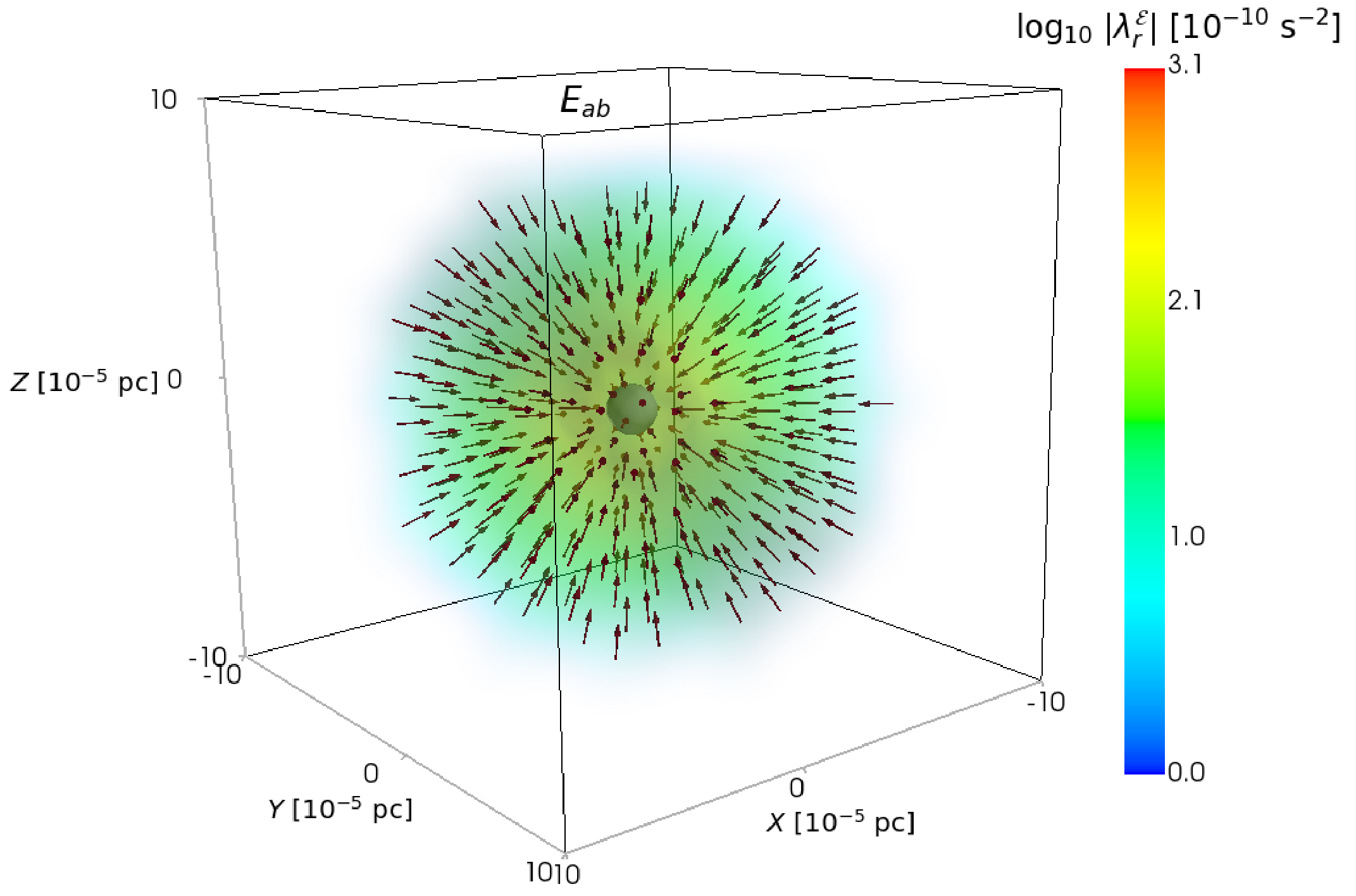}\\
\includegraphics[width=0.75\textwidth, trim = 0 0 0 0, clip, angle=0]{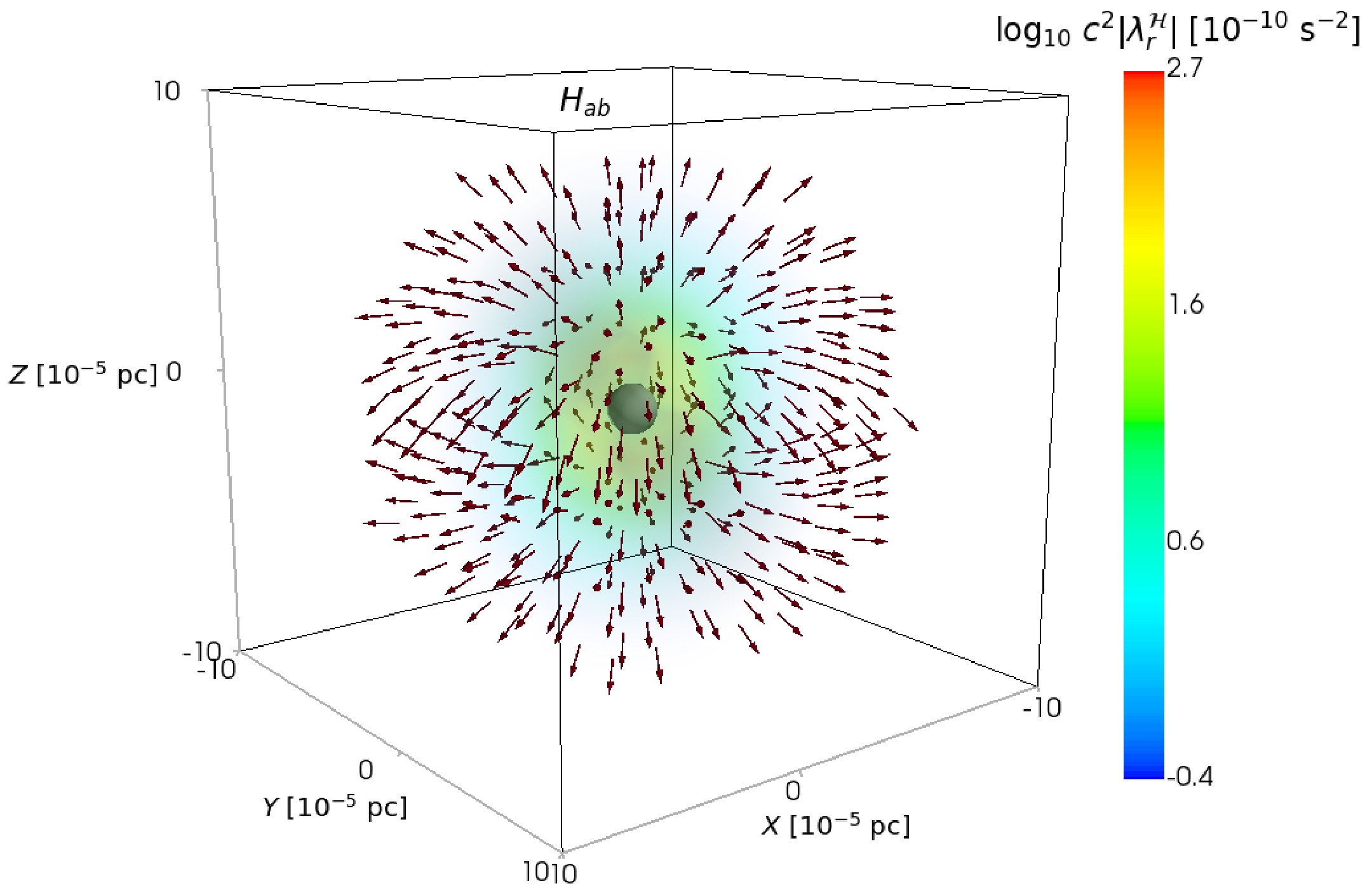}
\end{center}
\vspace{-5mm}
\caption{The gravitoelectric tensor field $E_{ab}$ (top panel) and  gravitomagnetic tensor field $H_{ab}$ (bottom panel) of a slowly rotating supermassive black hole with the dimensionless spin parameter of $a_{\ast} =0.5$ and mass of $M = 10^8 {\rm M}_{\odot}$. 
The color codes show the absolute value of the radial distance eigenvalues $\log_{10} |\lambda_{r}%
^{\mathcal{E}}|$ and $\log_{10} c^2|\lambda_{r}^{\mathcal{H}}|$, the vector arrows visualize the the radial distance eigenvectors regulated by the signs of the radial distance eigenvalues,  ${\rm sgn}(\lambda_{r}^{\mathcal{E}}) \vec{V}_{r}^{\mathcal{E}}$ and ${\rm sgn}(\lambda_{r}^{\mathcal{H}}) \vec{V}_{r}^{\mathcal{H}}$. 
\label{fig1}
}
\end{figure*}

In general relativity, we describe a black hole (BH) by three fundamental quantities: mass $M$, spin $a_{\ast}$ and charge $Q$.\cite{Chandrasekhar1983} The dimensionless spin parameter ($-1 \leq a_{\ast} \leq +1$) is defined as $a_{\ast} = Jc/GM^2$, where $J$ is the BH angular momentum, and $M$ is the BH mass. Negative values of $a_{\ast}$ describe retrograde rotation in which the black hole rotates in the opposite direction to the accretion disk, while positive values are associated with prograde rotation, and $a_{\ast} = 0$ implies no rotation. As a charged black hole would be rapidly neutralized by the accretion of oppositely charged particles, the charge quantity $Q$ could be  negligible. 
Let us consider a slow rotating BH described by Ref.~\citenum{Zhang2012} in the Boyer--Lindquist coordinates 
(radial distance: $r$, polar angle: $\theta$, azimuthal angle: $\varphi$).
In the Kerr metric, we use the Kerr spin parameter $a \equiv a_{\ast} GM /c^2$ with the dimension of length. 
We then obtain the gravitoelectric and gravitomagetic tensor fields for a
slow rotating SMBH with a spin parameter of $a_{\ast}=0.5$ and a mass of $M=10^8 {\rm M}_{\odot}$, and
solve $E_{ab}V_{\mathcal{E}}^{a}$ $=$ $\lambda_{\mathcal{E}}V_{b}%
^{\mathcal{E}}$ and $H_{ab}V_{\mathcal{H}}^{a}$ $=$ $\lambda_{\mathcal{H}%
}V_{b}^{\mathcal{H}}$ for the radial distance $r$, where $V_{a}^{\mathcal{E}}$
and $V_{a}^{\mathcal{H}}$ are the eigenvectors, and $\lambda^{\mathcal{E}}$
and $\lambda^{\mathcal{H}}$ are the eigenvalues of $E_{ab}$ and $H_{ab}$,
respectively. The absolute values of the radial distance eigenvalues $|\lambda_{r}%
^{\mathcal{E}}|$ and $|\lambda_{r}^{\mathcal{H}}|$ 
correspond to the amplitudes of the gravitoelectric tensor $E_{ab}$ and
gravitomagetic tensor $H_{ab}$ as functions of the radial distance $r$, respectively. 
The radial distance eigenvectors 
$\vec{V}_{r}^{\mathcal{E}}$ and $\vec{V}_{r}^{\mathcal{H}}$  
visualize the physical lines of the gravitoelectric and
gravitomagetic tensor fields.
Figure~\ref{fig1} shows  $\log_{10} |\lambda_{r}^{\mathcal{E}}|$ and $\log_{10} c^2|\lambda_{r}^{\mathcal{H}}|$ by color codes,  
and ${\rm sgn}(\lambda_{r}^{\mathcal{E}}) \vec{V}_{r}^{\mathcal{E}}$ and ${\rm sgn}(\lambda_{r}^{\mathcal{H}}) \vec{V}_{r}^{\mathcal{H}}$
by vector arrows (${\rm sgn}(x)$ is the signum function). The physical lines shown 
for ${\rm sgn}(\lambda_{r}^{\mathcal{H}}) \vec{V}_{r}^{\mathcal{H}}$ 
are comparable to the unified outflow model proposed for 
ultra-fast outflows 
observed 
in high-energy X-ray observations of 
active galactic nuclei.\cite{Kazanas2012,*Tombesi2013} 
In particular, the measurements of SMBH spins are now possible with the recent advancements in X-ray astronomy, 
\cite{Reynolds2019,*Reynolds2014,*Brenneman2013,Brenneman2006,*Miniutti2007,*Zoghbi2010,*deLaCallePerez2010,*Patrick2011,*Nardini2011,*Brenneman2011,*Tan2012,*Fabian2013,*Lohfink2013,*Walton2013} 
so we could examine whether SMBH angular momenta are correlated with outflow kinematics and density profiles.
From Figure~\ref{fig1}, it can be seen that $|\lambda_{r}^{\mathcal{H}}|$
has its maximum value at regions at the north and south poles outside the event horizon, while
it vanishes at the boundary of the event horizon, so
the gravitomagnetic field may support outflows of accreted materials along the BH spin axis far from the event horizon of the spinning BH.
This is in agreement with the Penrose mechanism \cite{Penrose1969,*Penrose2002,*Penrose1971} that explained how
rotational energy is extracted from a Kerr BH.

We might generalize the \textit{slow-Kerr} metric of Ref.~\citenum{Zhang2012} 
to a
perturbed FLRW spacetime that is applicable to a large supermassive non-compact
object slowly rotating in almost-FLRW model such as a massive disc galaxy.
To describe a galaxy, the BH spin parameter $a_{\ast}$ is replaced with a dimensionless spin
parameter $\lambda_{\ast}$, so we may define a perturbed FLRW
spacetime as follows: 
\begin{align}
ds^{2}=  &  -\left(  1+\frac{2\Phi(M,r,\theta)}{c^{2}}\right)  c^{2}%
dt^{2}+\left(  1+\frac{2\Phi(M,r,\theta)}{c^{2}}\right)  ^{-1}dr^{2}%
\nonumber\\
&  +r^{2}\left(  d\theta^{2}+\sin^{2}\theta d\varphi^{2}\right)
+\frac{4\lambda\Phi(M,r,\theta)}{c^{2}}\sin^{2}\theta cdtd\varphi,
\end{align}
where $\Phi(M, r,\theta)$ is the Newtonian gravitational potential, and $\lambda \equiv \lambda_{\ast} GM /c^2 $
is a parameter with the dimension of length that characterizes the rotation.

For disc-like galaxies, we may define the gravitational potential $\Phi$ based on the generalized Pulmmer\cite{Plummer1911}'s three-dimensional mass model in the spherical coordinates 
as follows \cite{Miyamoto1975,*Kuzmin1956} %
\begin{equation}
\Phi(M,r,\theta)=-\frac{GM}{ \big( r^2 +[R_{\rm a}+(R^2_{\rm b} + r^2\cos^2\theta )^{1/2}]^2 \big) ^{1/2}}, %
\end{equation}
and the dimensionless spin parameter $\lambda_{\ast}$ as \cite{Peebles1969,*Peebles1980,*Peebles1993} %
\begin{equation}
\lambda_{\ast}=\frac{J\left\vert E\right\vert ^{1/2}}{GM^{5/2}},%
\end{equation}
where $R_{\rm a}$ and $R_{\rm b}$ are constants with the dimension of length characterizing various non-spheroidal mass distributions, 
$J$ is the total angular momentum, $E$ is the total binding energy, and $M$ is the
total mass. 
The spin parameter is typically a low value around $\lambda_{\ast} \approx 0.05$ for elliptical
(\textit{non-disc}) galaxies, but a larger value about $\lambda_{\ast} \approx 0.5$ reported for spiral and lenticular (\textit{disc}) galaxies. \cite{Efstathiou1979,*Fall1980,*Kashlinsky1982,*Davies1983,*Barnes1987,*Warren1992,*Catelan1996}

In this configuration, the weak production of gravitomagentic fields on both sides of the galactic disc
might be expected by the rotation of a massive spiral galaxy typically having baryonic masses
of $10^{8.5}$--$10^{11.5}\mathrm{M}_{\odot}$ \cite{Salucci1999,*Papastergis2012,*Perez-Gonzalez2008,*Bernardi2010,*Baldry2012}   
and spins of $\lambda_{\ast} \approx 0.5$. \cite{Efstathiou1979,*Fall1980,*Kashlinsky1982,*Davies1983,*Barnes1987,*Warren1992,*Catelan1996} In the case of an \textit{active} galaxy 
containing a rapidly spinning SMBH at its center, we may also expect the strong production of
gravitomagentic fields near the galactic center along the spin axis powered by the spinning SMBH
typically having masses of 
$10^6$--$10^{9}\mathrm{M}_{\odot}$ \cite{Tonry1987,*Dressler1988,*Kormendy1988,*Kormendy1992,*Kormendy1996,*Kormendy1997,*Magorrian1998,*Cretton1999} 
and some having spins of $a_{\ast}\approx 0.9$ (measured in several active galaxies from relativistically broadened X-ray K$\alpha$ iron lines 
\cite{Brenneman2006,*Miniutti2007,*Zoghbi2010,*deLaCallePerez2010,*Patrick2011,*Nardini2011,*Brenneman2011,*Tan2012,*Fabian2013,*Lohfink2013,*Walton2013}). This phenomenon can be explained by Eq.~(\ref{eq2}) that
associates the gravitomagentic field production with the angular momentum density.
Other possible phenomena are predicted by Eqs.~(\ref{eq3}) and (\ref{eq4}) where
the curl (and temporal variation) of the gravitomagentic field
contributes to the temporal variation (and curl) of the gravitoelectric 
field. 

Can the rotation of a massive disc galaxy and its rapidly spinning SMBH contribute
to the production of gravitomagentic fields on both sides of the galactic disc and along the SMBH spin axis? Can
these gravitomagentic fields cycling over the galactic disc induce some
gravitoelectric fields into rotational motions of stars within the galactic disc? Implications of the conformal Weyl gravity for galactic rotation curves have been explored by Ref.~\citenum{Mannheim1989,*Mannheim1990,*Mannheim1992,*Mannheim1994,*Mannheim1997,*Mannheim2001,*Mannheim2012},
but using the Schwarzschild solution, which could not adequately explain discrepancies in rotational velocity curves between 
elliptical galaxies ($\lambda_{\ast} \approx 0.05$) and spiral (disc) galaxies ($\lambda_{\ast} \approx 0.5$).
It is worthwhile considering whether the current issues in rotation curves of disc galaxies could be resolved by the equations
of motion for the gravitoeletric and gravitomagentic fields in a perturbed FLRW model.

How could be the interaction
between two massive \textit{active} galaxies due to their weak
gravitomagentic fields on large scales? In particular, some recent $N$-body
computational simulations of a perturbed FLRW spacetime \cite{Adamek2016} imply that the frame-dragging vortex, which is
expected to be large on small scales (e.g. near SMBH), could be at smaller
orders but considerable on large scales, and be also enhanced as the
universe is evolving from the primeval at the redshift $z \sim10$ to
the present-day one at $z \sim0$.
We know that the universe just after cosmic reionization ($z\sim 6$) contained mostly low-mass \textit{starburst} dwarf galaxies,
which were gradually evolving into massive \textit{quiescent} and \textit{active} galaxies at cosmic noon ($z \sim 1.5$--$3$) 
due to multiple galaxy merger events.
We do not yet fully comprehend how this galaxy evolution influenced our universe on large scales.


%

\end{document}